\documentclass[sigconf,10pt,letterpaper,noacm]{acmart}
\AtBeginDocument{%
  \providecommand\BibTeX{{%
    \normalfont B\kern-0.5em{\scshape i\kern-0.25em b}\kern-0.8em\TeX}}}

\acmConference[]{}{}{}
\acmBooktitle{}
\acmYear{}
\copyrightyear{}
\acmDOI{}
\acmISBN{}
\usepackage{tablefootnote}

\usepackage{tabulary}
\usepackage{circuitikz}
\usepackage{caption}
\usepackage{subcaption}
\usepackage{xspace}
\usepackage[binary-units=true]{siunitx}
\usepackage{multirow}
\usepackage{marvosym}
\usepackage{hyperref}
\usepackage{siunitx}
\usepackage{diagbox}
\usepackage{makecell}
\usepackage{color, colortbl}
\usepackage{lipsum}
\usepackage{threeparttable}
\definecolor{LightRed}{rgb}{1, 0.6, 0.6}
\definecolor{LightGreen}{rgb}{0.0, 0.8, 0.6}
\definecolor{LightBlue}{rgb}{0.61, 0.87, 1.0}
\definecolor{Grey}{rgb}{0.66, 0.66, 0.66}

\usepackage{filecontents}
\usepackage{soul}
\usepackage{caption}
    \usepackage{subcaption}
    \usepackage{graphicx}
\usepackage{diagbox}
 \usepackage{float}
 \usepackage{tabulary}
\usepackage{circuitikz}
\usepackage{caption}
\usepackage{subcaption}
\usepackage{xspace}
\usepackage[binary-units=true]{siunitx}
\usepackage{multirow}
\usepackage{marvosym}
\usepackage{hyperref}
\usepackage{siunitx}
\usepackage{diagbox}
\usepackage{makecell}
\usepackage{color, colortbl}
\usepackage{lipsum}
\usepackage{threeparttable}
\usepackage{enumitem}
\usepackage{comment}
\usepackage{float}
\usepackage[toc,page]{appendix}
\usepackage[official]{eurosym}
\usepackage[english]{babel}
\usepackage{fancyhdr}
\usepackage{svg}

\usepackage{mhchem}

\begin{document}


\title{Photoacoustics on the go:\\An Embedded Photoacoustic Sensing Platform}




\setcopyright{none}
\settopmatter{printacmref=false}
\renewcommand\footnotetextcopyrightpermission[1]{}


\begin{abstract}
Several centimeters below the skin lie multiple biomarkers, such as glucose, oxygenation, and blood flow. Monitoring these biomarkers regularly and in a non-invasive manner would enable early insight into metabolic status and vascular health. 
Currently, there are only a handful of non-invasive monitoring systems.
Optical methods offer molecular specificity (i.e., multi-biomarker monitoring) but have shallow reach (a few millimeters); ultrasound penetrates deeper but lacks specificity; and MRI is large, slow, and costly. Photoacoustic (PA) sensing combines the best of optical and ultrasound methods. A laser transmitter emits pulses that are absorbed by different molecules, providing specificity. These light pulses generate pressure changes that are captured by an ultrasound receiver, providing depth. Photoacoustic sensing is promising, but the current platforms are bulky, complex, and costly.
\textit{We propose the first embedded PA platform}. Our contributions are fourfold. First, inspired by LiDAR technology, we propose a novel transmitter that emits pulses similar to those in the state-of-the-art (SoA), but instead of using high-voltage sources and complex electronic interfaces, we use a simple low-power microcontroller (MCU). Second, we carry out a thorough analysis of our custom transmitter and a commercial system. Third, we build a basic ultrasound receiver that is able to process the faint signal generated by our transmitter. Lastly, we compare the performance of our platform against a SoA commercial system, and show that we can detect glucose and (de)oxygenated hemoglobin in two controlled solution studies. The resulting signal characteristics indicate a plausible path toward noninvasive, real-time, at-home sensing relevant to diabetes care.
More broadly, this platform lays the groundwork for translating the promise of PA sensing into a broader practical reality.

\end{abstract}

\author{Talia Xu}
\email{talia.xu@auckland.ac.nz}
\affiliation{
  \institution{University of Auckland}
}

\author{Caitlin Smith}
\email{csmi310@aucklanduni.ac.nz}
\affiliation{
  \institution{University of Auckland}
}

\author{Charles Lo}
\email{charles@charleslo.net}
\affiliation{}

\author{Jami Shepherd}
\email{jami.shepherd@auckland.ac.nz}
\affiliation{
  \institution{University of Auckland}
}

\author{Gijs van Soest}
\email{g.vansoest@erasmusmc.nl}
\affiliation{
  \institution{Erasmus MC}
}

\author{Marco Zuniga}
\email{M.A.ZunigaZamalloa@tudelft.nl}
\affiliation{
  \institution{TU Delft}
}

\renewcommand{\shortauthors}{Xu et al.}

\maketitle

\vspace{-1 mm}
\section{Introduction}
\label{sec:1-introduction}

Monitoring the human body several centimeters underneath the skin in a low-cost and non-invasive way would represent a paradigm shift in physiological tracking. 
Such systems would allow measuring multiple biomarkers simultaneously, such as glucose, oxygen saturation, and blood flow. 
Real-time measurements of these biomarkers could reveal subtle physiological changes in the body long before they are visible clinically, allowing new potentials in monitoring health, treatment efficacy, and recovery \cite{Tayyari2015, Okamoto1998, Khoobehi2013, Feldman2019}. This capability remains out of reach for existing sensing technologies, such as MRI and ultrasound, but it is exactly the type of opportunity that photoacoustic (PA) sensing makes possible. Beyond human health, PA sensing also has applications in areas such as environmental monitoring, food quality control, and material analysis \cite{Schnaiter2023, Sheridan2005, Patimisco2014, Herrmann2021, Fiorani2021, Lv2018}.

 \begin{figure*}[h]
    \centering
    \begin{subfigure}[t]{0.24\linewidth}
    \includegraphics[width=0.963\linewidth]{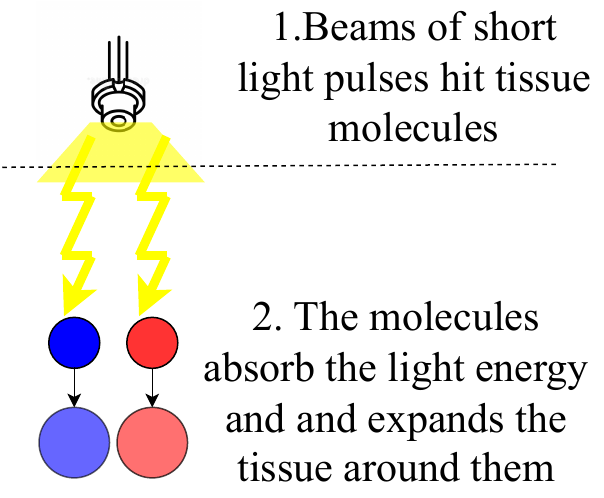}
    \caption{Excitation.} 
    \label{fig:PA-step-1}
    \end{subfigure} 
    \hspace{1.5mm}
    \begin{subfigure}[t]{0.33\linewidth}
    \includegraphics[width=\linewidth]{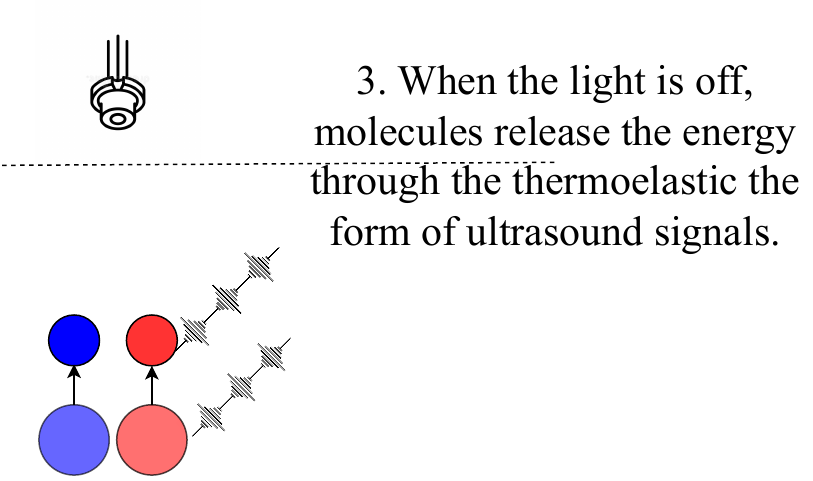} 
    \caption{Energy release.}
    \label{fig:PA-step-2}
    \end{subfigure} 
    \hspace{1.51mm}
    \begin{subfigure}[t]{0.33\linewidth}
    \includegraphics[width=\linewidth]{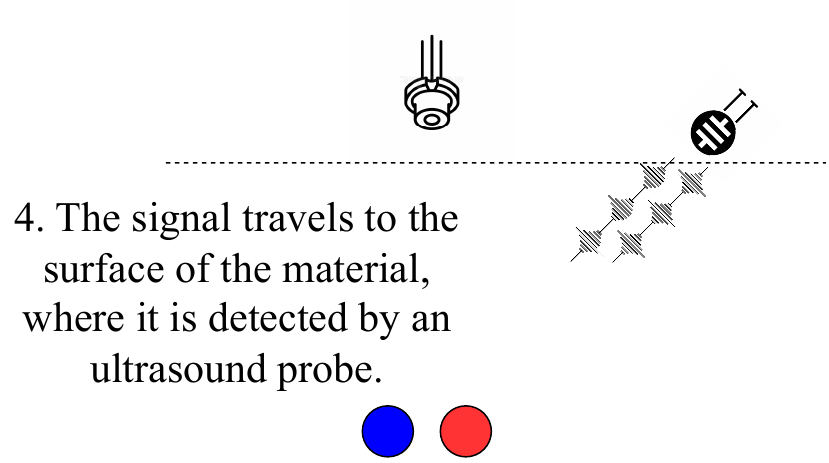} 
    \caption{Wave propagation.}
    \label{fig:PA-step-3}
    \end{subfigure} 
\caption{Key Steps in PA signal transmission and reception.}
\label{fig:PA-step}
\end{figure*}

Compared to technologies that use a single physical signal for monitoring, such as sound, light, or magnetic fields, PA uniquely combines the strength of two physical signals: sound and light, as illustrated in \autoref{fig:PA-step}. 
In the first stage, short pulses of laser light are sent into the target sample. When the molecules absorb the laser energy, the medium around them undergoes a rapid thermal expansion. During the “off” period between pulses, the tissue releases that energy as low-amplitude compressional waves that can be detected with ultrasound sensors.  

PA sensing offers advantages over other techniques. Purely optical methods can identify molecules but are limited to shallow depths (1~mm to 2~mm compared to up to 5~cm in PA \cite{Lin2022NRCO_PAI_Oncology}). Ultrasound can reach deeper but sees only structure and is sensitive to mechanical variations, which leads to weak specificity in biological tissues. Large systems like MRI can provide detailed anatomy, but they are slow and expensive, and cannot directly
track molecular or functional parameters at the fine-scale PA can. Furthermore, MRI is not sensitive to endogenous molecules such as hemoglobin without external contrast agents. 

In PA sensing, the signal strength is proportional to the concentration of target molecules. The more molecules that absorb the light, the stronger the resulting ultrasound signal. This quantitative link between molecular concentration and acoustic response enables not only the identification of biomarkers but also the estimation of their concentration. The ability to combine depth, specificity, and real-time measurement opens opportunities far beyond current devices, from biomedical monitoring to applications in materials inspection, food quality assessment, and environmental sensing.

Current commercial PA systems hold broad promise across many fields, but their use is still largely confined to research settings. 
They are bulky and non-portable, relying on high-power lasers that require advanced cooling and strict safety protocols. Furthermore, the systems require high sampling rates, long acquisition times, and intensive computational processing.
These limitations in size, complexity, and operation have so far kept PA out of field-deployable devices. 

In this work, we take the first step toward breaking these barriers. We leverage advances made by LiDAR technology in compact, low-cost pulsed lasers to demonstrate core PA sensing functionality in a smaller and more practical form. Our design includes a novel transmitter, receiver, benchmarking against commercial systems, and a preliminary evaluation. In particular, our contribution is fourfold.

\textit{\textbf{Contribution 1 [\autoref{sec:3-transmitter}]} Demonstrating the feasibility of an embedded PA transmitter.}: One of the central unknowns in bringing PA into embedded systems is whether a low-power, compact transmitter can generate signals strong enough to be detected. We design and implement a novel embedded PA transmitter board that integrates low-cost laser diodes with an efficient driver and control managed by a small on-board microcontroller.

\textit{\textbf{Contribution 2 [\autoref{sec:4-watertank}]}: Comparative analysis with a commercial PA system.} We characterize the output of our transmitter and show that, despite having significantly lower pulse energy than lab-scale systems, the transmitter can still produce measurable signals comparable to commercial systems. To understand the practical trade-offs between embedded and commercial PA systems, we perform side-by-side measurements in a controlled water tank environment.

\textit{\textbf{Contribution 3 [\autoref{sec:5-receiver}]}: Embedded PA receiver with on-board processing.} We design a compact receiver board with integrated amplification, filtering, digitization, and on-board processing on an FPGA. This removes the need for bulky, external data acquisition systems and demonstrates that PA signal capture and initial processing can be achieved within the constraints of an embedded platform.

\textit{\textbf{Contribution 4 [\autoref{sec:6-evaluation}]}: Proof-of-principle biomarker detection.} We validate our system by detecting two representative biomarkers, glucose concentration and blood oxygenation analogues, in controlled solutions. 
These results point towards a future where the unique capabilities of PA sensing can transition from specialized laboratory equipment into compact, embedded systems, enabling continuous, multi-biomarker sensing.
\vspace{-1 mm}
\section{Background}
\label{sec:2-background}

\subsection{Principles of PA Sensing}

\begin{figure}[t]
    \centering
    \includegraphics[width=0.88\linewidth]{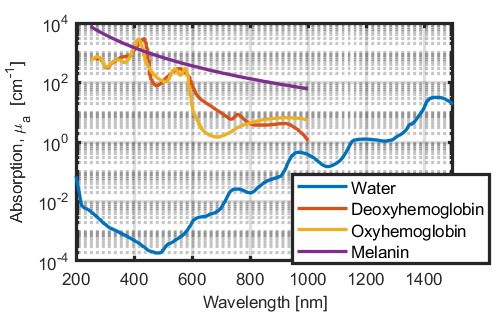}
    \vspace{-1mm}
    \caption{Absorption coefficient of water, oxy-hemoglobin and deoxy-hemoglobin with wavelength \cite{Segelstein1981WaterIndex, PrahlGratzerKollias_HbExtinction_OMLC, Jacques1996MelanosomesSPIE, Jacques1991MelanosomeThreshold}.}
    \vspace{-3mm}
    \label{fig:molecule-absorption}
\end{figure}

PA sensing uses light-induced acoustic signals to probe a sample. A laser pulse (“on”) deposits energy that produces a thermoelastic expansion. When the pulse ends (“off”), the expansion relaxes and emits an ultrasonic wave that can be detected with an ultrasound transducer. 
Target molecules under the skin absorb the laser's light selectively. For example, as shown in \autoref{fig:molecule-absorption}, water strongly absorbs energy in some specific bands, which are different from hemoglobin. For example, at short wavelengths (approximately $200$--$800\ \mathrm{nm}$), water's absorption coefficient is orders of magnitude lower than that of hemoglobin or melanin.

The process begins when a nanosecond-length laser pulse, typically less than 100~ns, rapidly deposits energy into a material. The short pulse length is crucial because it ensures that during illumination, both the heat diffusion and mechanical expansion are negligible, effectively confining the thermal and stress energy to the localized area until the pulse ends. This confinement creates a sharp, well-defined acoustic signal. The initial pressure, $p_0$, generated at the location of absorption is the fundamental signal in PA sensing. Its amplitude is directly proportional to the absorbed optical energy and can be described by the following equation:

\begin{equation}
    p_0(\mathbf{x}) = \Gamma\,\mu_a(\mathbf{x})\,F,
    \label{eq:PA-pressure}
\end{equation}

where \textbf{$p_0(\mathbf{x})$} is the initial pressure rise at a specific location $\mathbf{x}$, which is directly related to the detected ultrasound signal. This pressure is the product of three factors: \textbf{$\Gamma$} describes how efficiently heat is converted into pressure; \textbf{$\mu_a(\mathbf{x})$}, the optical absorption coefficient, which is the critical sensing parameter as it is proportional to the molecule's concentration; and \textbf{$F$}, the optical fluence, which represents the laser energy delivered per unit area.

\vspace{-2mm}
\subsection{System Design Considerations}
\autoref{eq:PA-pressure} provides a key insight for our system design: the measured acoustic signal can be directly used to quantify the target molecule's concentration. This also establishes specific design objectives for building effective PA sensing systems. 
\textbf{\textit{For the transmitter}}, a fundamental requirement for generating a PA signal is to create a rapid temperature change in the sample. The most direct method to achieve this is by delivering light in brief, high-energy \textit{pulses}, which is the focus of this design. However, it's also possible to generate PA signals by rapidly modulating the intensity of a continuous wave (CW) laser. Although intensity-modulated CW sources can produce PA signals, achieving the required modulation depth and power is more complex. Therefore, in this work we focus on pulsed lasers.
We also need multiple wavelengths to differentiate the various types of absorbing molecules.
The pulses generated by these lasers have three important design parameters: pulse energy, pulse width, and pulse repetition rate. 
These parameters are directly controlled by the system design and are critical for dictating the optical fluence delivered to the sample and the resulting PA signal properties. In the next section, we analyze these three parameters in detail.



\textbf{\textit{For the receiver}}, an ultrasound transducer converts the acoustic pressure waves generated within the material into an electrical signal.
The transducer is sensitive to mechanical energy and, in the context of PA sensing, function to capture the transient pressure changes caused by optical absorption. 
A core challenge is to design the electronic and processing capabilities to amplify small PA signals without introducing substantial noise, while also supporting high-speed data acquisition and efficient on-device processing. 



\textbf{\textit{For the material under test,}} the key principle of PA sensing is that different molecules absorb specific colors (wavelengths) of light in a unique way, like a 'color fingerprint'. This property enables two main types of analysis. For a foundational task like tracking the concentration of a single substance, one carefully chosen wavelength can be sufficient. 
However, for more advanced applications, such as differentiating multiple substances within a mixture, using multiple wavelengths is beneficial. Varying the illuminating wavelength can selectively excite different substances within the material, enabling their identification and quantification based on their unique PA responses.
To validate these capabilities, it's crucial to test the system on simple, controlled solutions before moving to complex biological tissues. Therefore, this paper focuses on this foundational validation using such solutions. 
\vspace{-2mm}
\subsection{Safety Considerations}

Commercial PA systems use high-brightness, nanosecond class-IV lasers, which raises concerns for both eye safety and skin exposure limits. The maximum permissible exposure (MPE) depends on optical wavelength, pulse width, beam size at the tissue, and exposure time \cite{ICNIRP2013LaserMPE}. For repeated pulses there are two guidelines: (1) no single pulse may exceed the per-pulse MPE, and (2) the total energy in any time window must stay below the time-averaged MPE\footnote{For skin, typical nanosecond limits are $20~\mathrm{mJ/cm^2}$ in the visible spectrum (400–700~nm), rising to $\sim50~\mathrm{mJ/cm^2}$ at $905~\mathrm{nm}$ and $100~\mathrm{mJ/cm^2}$ for 1050–1400~nm.} 
Commercial systems often operate near these skin limits (tens of $\mathrm{mJ/cm^2}$ per pulse) to maximize SNR, which is why they also requires strict eye protection\footnote{The per-pulse \emph{eye} MPE at $905~\mathrm{nm}$ is on the order of $5\times 10^{-4}~\mathrm{mJ/cm^2}$, orders of magnitude below the skin limits \cite{ICNIRP2013LaserMPE}}. Thus, preventing any eye exposure with barriers and eyewear is mandatory, making operation in open environments cumbersome.

Our compact design, on the other hand, minimizes exposure at the source. It uses sub-microjoule, tens-of-nanoseconds pulses in the visible and near-infrared, spreads or diffuses the beam to lower energy density, and enforces hard limits on per-pulse energy, repetition rate, and duty cycle. In a representative case ($905~\mathrm{nm}$, $220~\mathrm{nJ}$ at $10~\mathrm{kHz}$, unfocused at $\sim 1~\mathrm{cm}$), the \emph{skin} fluence per pulse is $\sim 1.4\times 10^{-3}~\mathrm{mJ/cm^2}$, about $10^{4}$–$10^{5}\times$ below the skin MPE. The 10~s average irradiance is $\sim 7~\mathrm{mW/cm^2}$, far below the $\sim500~\mathrm{mW/cm^2}$ limit at 905~nm. Eye risk can be further minimized by incorporating the source in a wearable design because the emitter works in contact mode with mechanical baffling and enclosures that remove any line of sight. 
\vspace{-1 mm}
\section{Transmitter Design}
\label{sec:3-transmitter}

In \autoref{sec:2-background}, we identified three critical parameters for the pulses generated by PA transmitters: pulse width, power per pulse, and repetition rate. The \textit{width} of the pulse needs to be short, less than 200 nanoseconds, to confine heat to a small region and produce a sharp acoustic response. A minimum amount of \textit{power per pulse} is essential to generate a detectable signal at the receiver. And the \textit{repetition rate} allows sending multiple subsequent pulses at a high frequency, increasing the signal-to-noise ratio (SNR). 
Laboratory systems and state-of-the-art (SoA) studies meet these requirements~\cite{Riksen2023}, but at the cost of bulky optics, complex electronics, cooling hardware, and costs in the range of tens to hundreds of thousands of dollars. Our key research question is: \textit{Can we design a photoacoustic system controlled by a low-cost embedded platform?}

The different values that the width, pulse power, and repetition rate can take enable a flexible design space. To radiate the necessary energy over a tissue, PA systems have different design alternatives.  One option is to send a high pulse power (30 mJ), but at a slower repetition rate (10 Hz); while another design could use the opposite tradeoff, send pulses with 100 times lower pulse power but at a repetition rate that is 10{,}000 times higher~\cite{Erfanzadeh2019LowCostPAReview}. 

The challenge comes when we try to meet these same requirements in a compact, embedded form. With complex equipment, the entire design space is available because the transmitter can generate a wide range of different pulses. Embedded platforms, however, have limited energy, computational, and speed capabilities. Thus, two important questions need to be answered to enable embedded PA: \textit{What combinations of width, pulse power, and repetition rate are feasible with embedded systems?} And, \textit{are those pulses sufficient to trigger the required PA phenomenon?} 

\textit{The key contribution of our work is to bridge the gap between PA and embedded systems by building upon concepts from LiDAR technologies.} In this section, we present our proof-of-concept transmitter designed to bridge that gap. 

\vspace{-1mm}
\subsection{Laser Driver Circuit}

Commercial systems and SOA studies have (1) advanced signal generator capabilities, enabling the transmission of narrow pulses; and (2) high voltage sources, above 200\,V, enabling pulses with high pulse power. Embedded platforms do not have these capabilities. The slower processor constrains the pulse width and repetition rate, and the low voltage constrains the pulse power.

To attain the required narrow pulses for PA, our laser driver design draws on methods developed in Light Detection and Ranging (LiDAR) applications, where laser diodes are driven in short pulses to enable fine ranging resolutions. 
However, LiDAR applications have very different design considerations. For example, they focus on timing accuracy rather than high optical power, since eye-safety regulations in an open environment prevent strong light emission. 

\begin{figure}
    \centering
    \includegraphics[width=0.5\linewidth]{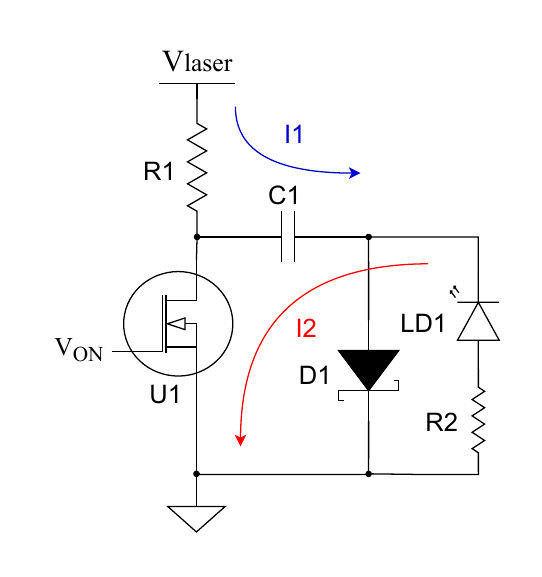}
    \caption{PA Driver Circuit}
    \label{fig:PA-driver-circuit}
\end{figure}

The challenge is that, to excite PA signals, we need a laser pulse that is as short as in LiDAR but also stronger at the same time. Our key insight to enable the required \textit{pulse width} and \textit{pulse power} for embedded PA is to adapt the two-stage structure from LiDAR drivers to support stronger discharge currents in a compact and embedded form, as shown in \autoref{fig:PA-driver-circuit}. To understand our design, we need to describe the two-stage LiDAR approach. 

\textbf{The first stage} is a slow charging path, in the order of a few $\mu$s (\(V_{\text{laser}}\)-R1-C1). A power source ($V_{laser}$) charges a capacitor ($C1$) to a set voltage while the laser diode ($LD1$) is disconnected from the circuit path. This stage defines the energy available for each pulse and keeps supply noise away from the diode. 
The size of the capacitor determines a lower bound for the pulse width and an upper bound for the repetition rate, which are central for PA. The energy stored in the capacitor is proportional to the voltage and capacitance that it can store according to the well-known equation:

\begin{equation}
    E = \frac{1}{2} C V^2,
    \label{eq:laser-ev}
\end{equation}

where $C$ is the capacitance and $V$ is the voltage the capacitor charges to. This means that the pulse energy can be increased either by increasing the charge voltage, or by increasing the capacitance.  On embedded, low-power platforms, the voltage $V$ is often constrained. On the other hand, increasing the capacitance $C$ lengthens the recharge time (reducing the maximum repetition rate) and increases the pulse width.

\textbf{The second stage} is the fast path (U1-C1-LD1-R2). In an ideal case, all of the stored energy would flow into the laser diode instantaneously, producing an impulse-like current. In reality, parasitic inductance and resistance, along with the finite switching speed of the device, limit how quickly this transfer can happen. These effects distort the pulse and increase its width, making circuit layout and component choice critical. 

Overall, considering the two stages, our design focuses on two main questions. First, what circuit design choices would limit (dis)charge times and reduce pulse distortions? Second, what are the maximum voltage and capacitance that a low-power design should have to generate PA pulses?  We answer these questions in this section. 
The division into slow and fast paths separates energy storage from energy delivery, allowing us to explore pulse width, power, jitter, and diode stress in a controlled way.  
With our approach, a design pattern that originated in LiDAR for short, eye-safe pulses becomes the basis for an integrated, higher-power driver suited to PA transmission.

\vspace{-1mm}
\subsection{Laser Diode Selection}

The vast majority of commodity laser diodes are optimized for low-cost, continuous-wave (CW) operation for mass-market applications like communications and illumination.
For generating high-current, nanosecond pulses needed for PA signals, specialized pulsed laser diodes (PLDs) are required, but they are expensive and have limited availability.
This means that researchers often have to adapt commodity CW laser diodes instead.


Prior work has shown that CW diodes can be repurposed successfully to generate PA signals \cite{Seeger2023, Stylogiannis2018, Stylogiannis2022}. Seeger et al. showed that these CW diodes, in pulsed operation, can be driven well beyond their datasheet limits. They tolerated peak currents up to 44 A and repetition rates up to 600 kHz for several hours without damage \cite{Seeger2023}. These results establish a foundation for our work: low-cost diodes can serve as practical transmitters when driven by complex drivers and high-voltage sources. However, can those diodes still generate a PA signal with a low power, simpler embedded platform?

Building upon the SOA, we select laser diodes that are inexpensive and easy to source, but to consider a more comprehensive evaluation, we also chose wavelengths spanning the visible red (638 nm), near-infrared (904 nm), and deeper infrared (1470 nm). 
This allows us to test across devices with different optical characteristics, while also covering a spectrum from visible to NIR that is broadly relevant for sensing and biomedical applications.

\begin{table}[t]
\centering
\caption{Laser diode models and power rating.}
\vspace{-2mm}
\label{tab:laser-diodes}
\footnotesize
\begin{tabular}{c|c|c|c}
\hline
\textbf{Model} & \textbf{Wavelength} & \textbf{Power Rating} & \textbf{Type} \\
\hline
HL63373HD      & 638 nm  & 1 W   & CW \\ \hline
SPL90          & 905 nm  & 75 W  & Pulsed \\ \hline
1470A-96-1.3-1 & 1470 nm & 1.3 W & CW \\
\hline
\end{tabular}
\end{table}

\vspace{-1mm}
\subsection{Embedded Low-Power Control Circuit}

The limitations of current PA systems are that the drivers require high-voltage sources to attain a high \textit{pulse power} and high-speed logic to generate pulses with nanosecond \textit{pulse width}. On the other hand, an embedded PA platform must be portable, relying on a simple MCU and a low-power supply.

\textbf{A decoupled approach:} \textit{To obtain a narrow width.} Our key design choice is to decouple the control of the pulse \textit{width} from the MCU timing. 
The processor does not need to match the laser’s pulse length, unlike what is done in the SoA with high-speed processors.
Instead, the MCU only provides a trigger, and the width is set entirely by the capacitor discharge in hardware. In this way, we do not need specialized high-speed logic with nanosecond granularity; a low-cost, general-purpose MCU is sufficient to provide a trigger. This separation makes the design simpler and more flexible.

\textbf{A voltage and capacitance trade off:} \textit{To obtain the necessary pulse power and repetition rate.} An embedded platform does not provide high voltages, limiting the ability to generate a higher pulse power. This means that, even if we can send narrow pulses with our decoupled approach, the amount of energy reaching the tissue may not be sufficient to generate a recoverable PA signal.
To balance simplicity and higher signal energy, we restrict the input to a single low-voltage rail at 5V, removing dependencies on multiple power supplies. All additional voltage levels are generated on board through voltage regulators. Later, we show that with voltages in the range of 25-30\,V and capacitances in the range of 20-100\,nF, we can attain pulse powers of a few hundred nJ and repetition rates of tens of kilohertz, allowing our embedded platform to generate recognizable PA signals.

\begin{figure}[t]
    \centering
    \includegraphics[width=0.75\linewidth]{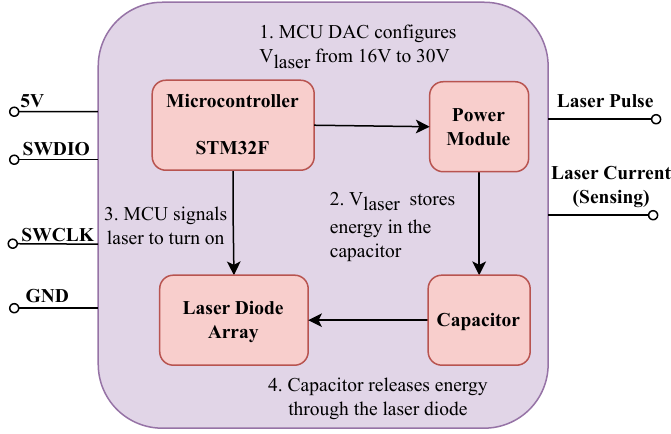}
    \caption{PCB Modules and control flow}
    \label{fig:PA-PCB-control}
\end{figure}

\textbf{Control modules.} \autoref{fig:PA-PCB-control} shows the overall control flow. Our platform relies on an STM32F303CBT6 MCU to manage timing, supply voltage, and monitor the system. 

\textit{Step 1:} On power‐up, the MCU uses its DAC to set the laser supply \(V_{\mathrm{laser}}\) (16–30 V).

\textit{Step 2:} A storage capacitor then charges to this $V_{laser}$ level, isolating the high-current discharge from supply noise, as discussed previously. 

\textit{Step 3:} The MCU outputs a short digital pulse that turns on the switch, discharging the capacitor through the diode. This digital pulse also acts as a sync signal for the receiver.

\textit{Step 4:} The capacitor releases the energy to the diode. A sensing resistor is used to measure the current through the diode, providing information for calibration, measurement, and feedback in later revisions.

\vspace{-1mm}
\subsection{Transmitter Prototype}

\begin{figure}[t]
    \centering
    \begin{subfigure}[t]{0.48\linewidth}
    \centering
    \includegraphics[width=0.9\linewidth, height=0.9\linewidth]{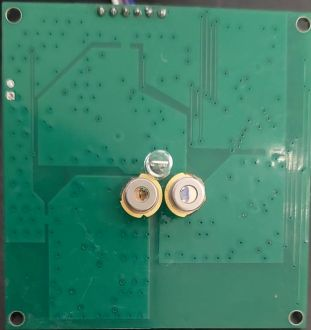}
    \caption{Front of PA.} 
    \label{fig:PA-prototype-front}
    \end{subfigure} 
    \hfill
    \begin{subfigure}[t]{0.48\linewidth}
    \centering
    \includegraphics[width=0.9\linewidth, height=0.9\linewidth]{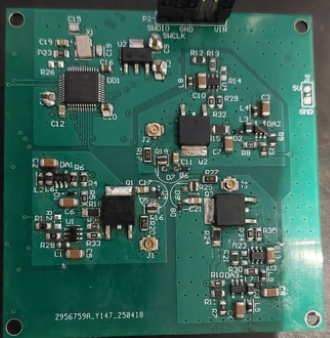} 
    \caption{Back of PA.}
    \label{fig:PA-prototype-back}
    \end{subfigure} 
\caption{PA prototype circuit with three laser diodes.}
\label{fig:PA-prototype}
\end{figure}

To demonstrate that compact and low-cost hardware can deliver reproducible nanosecond pulses for PA sensing, we developed a prototype board shown in \autoref{fig:PA-prototype}. The board has three laser diodes. The MCU sets the repetition rate in software, scalable into a few tens of kilohertz, so sensing speed and averaging can be adjusted without additional timing hardware. The pulse width is determined solely by the parasitics on the board (such as trace lengths), independent of the trigger input, which means the system only requires a slow digital trigger. As explained earlier, an important advantage of our design is that this separation also prevents timing jitter from translating into optical variation and keeps the output stable across bursts. \footnote{Another important design decision for the second stage is to keep the loop short to minimize inductance, the switch must operate with fast transitions to preserve edge speed, and clamp components need to suppress overshoot before it reaches the diode. Together, these constraints determine how closely the circuit can approach the ideal pulse and set the limits on rise time and peak current.}

\begin{figure}
    \centering
    \includegraphics[width=0.9\linewidth]{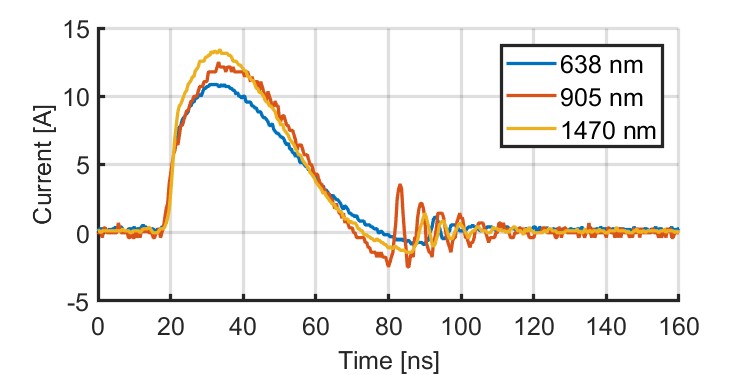}
    \caption{Current for three diodes in a single laser pulse.}
    \label{fig:voltage-laser-diode}
\end{figure}

\autoref{fig:voltage-laser-diode} shows the measured laser-diode current for a representative pulse. The waveform confirms a nanosecond-scale (around 40 ns) pulse shaped by the driver. In a 10-second run at 10 kHz (100,000 pulses), the peak current varies by less than 1\% across pulses. This validates that the circuit consistently converts a simple 5 V input into controlled, high-current pulses.
The measured pulse energies are consistent with those reported in prior work with more complex PA systems~\cite{Stylogiannis2018}, 
which indicates that even with reduced absolute energy, our custom driver reaches an operational range sufficient for biomedical use, as we showcase later on in our evaluation.

\begin{figure*}[t]
    \centering
    \begin{subfigure}[b]{0.28\linewidth}
        \centering
        \includegraphics[width=1\linewidth]{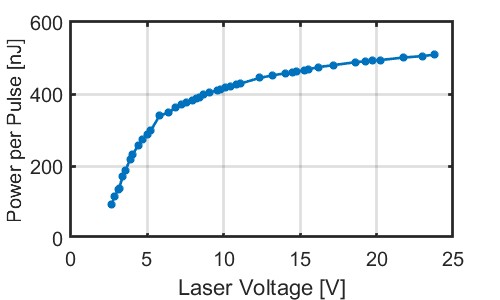}
        \caption{Pulse power with voltage.}
        \label{fig:ld-pulse-power-voltage}
    \end{subfigure}
    \hfill
    \begin{subfigure}[b]{0.28\linewidth}  
        \centering
        \includegraphics[width=1\linewidth]{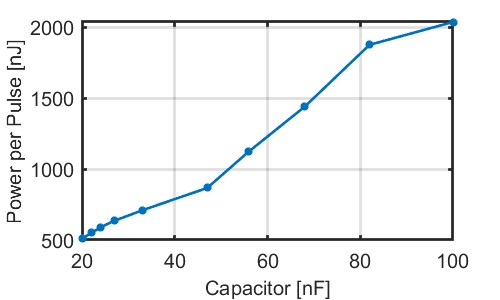}
        \caption{Pulse power with capacitance.}
        \label{fig:ld-pulse-power-cap}
    \end{subfigure}
    \hfill
    \begin{subfigure}[b]{0.3\linewidth}
        \centering
        \includegraphics[width=1\linewidth]{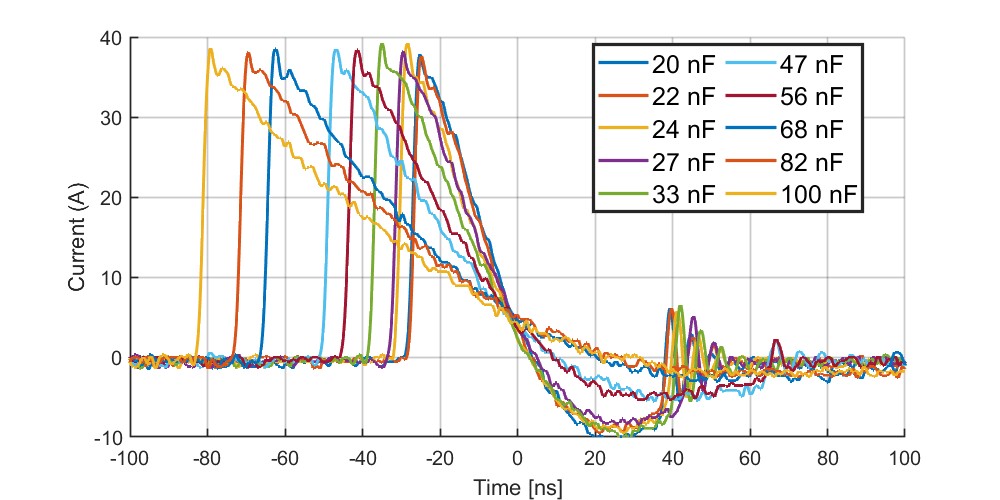}
        \caption{Pulse current with capacitance.}
        \label{fig:ld-pulse-width-cap}
    \end{subfigure}
\caption{PA power with varying voltage and capacitance for 638 nm laser diode HL63373HD (C1 in \autoref{fig:PA-driver-circuit})}
\label{fig:pa-pulse-varying}
\end{figure*}

\begin{table}[t]
  \centering
  \footnotesize
  \setlength{\tabcolsep}{12 pt}
  \begin{threeparttable}
  \vspace{3mm}
  \caption{Pulse energy comparison between the custom board and the commercial Radiant 532 LD system.}
  \vspace{-4mm}
  \label{tab:pulse-energy-comparison}
  \begin{tabular}{c|c|c}
    \hline
    \textbf{Wavelength} & \textbf{Custom PA Board} & \textbf{Radiant 532 LD}\tnote{a} \\
    \hline
    638 nm  & 520 nJ  & 180 $\upmu$J\tnote{b} \\
    \hline
    905 nm  & 2100 nJ & 115 $\upmu$J \\
    \hline
    1470 nm & 220 nJ  & 100 $\upmu$J \\
    \hline
  \end{tabular}

  \begin{tablenotes}
    \footnotesize
    \item[a] Lowest power output level of the system. Measured with an optical power meter.
    \item[b] Lowest wavelength of the Radiant 532 LD is 680 nm. This power is measured at 680 nm.
  \end{tablenotes}
  \end{threeparttable}
  \vspace{-2mm}
\end{table}

To place the prototype in context, we compare it against a commercial laser platform commonly used in PA research, the Opotek Radiant 532 LD. \autoref{tab:PA-comparison} summarizes the main differences. The prototype board is small and lightweight, while the Radiant is a large benchtop system weighing tens of kilograms. The cost difference is also substantial, with off-the-shelf diodes available for tens of dollars compared to roughly 100,000 USD for the commercial unit. 

The performance profiles reflect these differences in scale. The Radiant is capable of producing pulses with mJ energies and sub-10 ns widths, which remain beyond the capability of compact diode drivers. On the other hand, our prototype operates at repetition rates up to 100 kHz, much higher than the 10 Hz of the commercial system\footnote{The commercial system has the resources to generate repetition rates higher than ours, but they do not need to because thanks to their high voltages, individual pulses have a very high pulse power (more than 50 times higher than ours), requiring a much lower repetition rate to deliver the same energy on the tissue.}. This makes it particularly suited to applications that benefit from high averaging rates or fast scanning. These comparisons show that the two approaches could serve complementary roles. High-energy, tunable sources remain necessary where maximum optical output is required, while compact diode drivers enable low-cost, portable experiments that can be deployed outside traditional laboratory environments.

\begin{table}[h!]
    \centering
    \footnotesize
    \vspace{1mm}
    \caption{Comparison of the custom-designed PA board with the commercial Opotek Radiant 532 LD.}
    \vspace{-2mm}
    \label{tab:PA-comparison}
    \setlength{\tabcolsep}{2 pt}
    \begin{tabular}{c|c|c}
        \hline
        \textbf{Feature} & \textbf{Custom PA Board} & \textbf{Radiant 532 LD} \\
        \hline
        Size & 5 cm x 5 cm x 1 cm & 70 cm x 42 cm x 48 cm  \\
        \hline
        Weight & 50 g & 68 kg\\
        \hline
        \begin{tabular}[c]{@{}c@{}}Laser Range\end{tabular} & 638 nm, 905 nm, 1470 nm & Tunable 680 - 2600 nm\\
        \hline
        Pulse Width & 25 - 50 ns & 5 ns\\
        \hline
        Cost & \begin{tabular}[t]{@{}c@{}} 638nm: $\sim$\$15 USD \\  904nm: $\sim$\$7 USD \\  1470nm: $\sim$\$50 USD \\ Other: $\sim$\$20 USD \end{tabular} & $\sim$\$100,000 USD \\
        \hline
        Power & $\mu$J-level  & mJ-level \\
        \hline
        \begin{tabular}[c]{@{}c@{}}Repetition Rate\end{tabular} & Programmable, up to 100 kHz  & 10 Hz \\
        \hline
    \end{tabular}
    \vspace{-2mm}
\end{table}

\noindent\textit{Power consumption:} The prototype draws \(61 \mathrm{mA}\) from a \(5\,\mathrm{V}\) supply (\(\approx 305 \mathrm{mW}\)) while sequencing the three diodes at \(638\,\mathrm{nm}\), \(905\,\mathrm{nm}\), and \(1470\,\mathrm{nm}\) at \(15\,\mathrm{kHz}\). Because each pulse is around 50 ns (0.05 \% duty cycle at 10 kHz), the time-averaged electrical power delivered to the diodes is small. Accordingly, we measure essentially the same average power consumption whether operating a single diode or sequencing all three. The power budget is dominated by the MCU and the DC–DC converters rather than the laser pulses.

This approach differs fundamentally from high–energy commercial systems, which prioritize per–pulse optical energy and narrow bandwidth at the expense of wall–plug efficiency, often as low as 0.1\%, so the electrical burden grows quickly \cite{paschotta_wall_plug_efficiency_rp_photonics}. At \(0.1\%\) efficiency, delivering \(1\,\mathrm{W}\) of average optical power requires \(1{,}000\,\mathrm{W}\) of electrical input.

At our operating point, the system consumes \(\approx 0.6\,\mathrm{W}\) and produces \(\approx 14\,\mathrm{mW}\) of average optical power in total \((10.6\,\mathrm{mW}\ @\,905\,\mathrm{nm},\ 2.6\,\mathrm{mW}\ @\,638\,\mathrm{nm},\ 1.2\,\mathrm{mW}\ @\,1470\,\mathrm{nm})\), corresponding to a wall–plug efficiency of \(\approx 2.3\%\). Equivalently, achieving \(1\,\mathrm{W}\) of average optical power at this efficiency would require \(\approx 43\,\mathrm{W}\) of electrical input—about \(23\times\) less input power than a \(0.1\%\) - efficient system (and \(\sim 2.3\times\) less than a \(1\%\) system). These figures place our prototype in the complementary regime of low average power and compact implementation.
In addition, because the MCU (around 150 mW) and power electronics (\(generating V_{\text{laser}}\)) dominate the power budget, we can further reduce the power consumption by powering them down whenever the system is idle, and selecting a low-power MCU, since the design imposes no stringent GPIO- or clock-speed requirements.

\vspace{-1mm}

\subsection{Pushing the Limits of the Laser Diode}

With our laser diode transmitter, we generate much lower per-pulse energy than commercial systems, but at a much higher repetition rate. Because the signal strength of the transmitter is crucial to our signal quality, we need to also understand the limit of this architecture: how much energy can we deliver while preserving short pulses?

In this section, we bypass the on-board DC--DC converter and drive the laser charge node from an external supply, allowing a wide sweep of the charge voltage \(V_{\text{laser}}\) while independently varying the storage capacitance \(C_1\). These two parameters set the per-pulse energy, as shown in \autoref{eq:laser-ev}, and they also shape the optical waveform. Raising \(V_{\text{laser}}\) is the most direct way to increase energy without inherently lengthening the pulse, but it also increases electrical stress on the switch and laser diode and can introduce ringing and overshoot in the drive network. Increasing \(C_1\) also increases energy, but it extends recharge time and tends to broaden the pulse. If thermal or stress confinement conditions are not met (due to the pulse width being too large), the SNR of the PA signal degrades further.

As shown in \autoref{fig:pa-pulse-varying}, increasing the driver voltage raises per-pulse energy, but it also pushes the small, low-power laser diodes toward their limit. In our setup, about \(45\,\mathrm{A}\) peak current (roughly \(26\)–\(30\,\mathrm{V}\) supply) is already close to the maximum limit these laser diodes can tolerate. HL63373HD and HL63193MG at \(638\,\mathrm{nm}\), OPT1470 at \(1470\,\mathrm{nm}\) (all \(\sim 1\,\mathrm{W}\) CW), and ML562G86 (\(4.5\,\mathrm{W}\) CW) show similar breaking points clustered near \(28\)–\(30\,\mathrm{V}\). This indicates that our nominal operating point is already near the safe ceiling for the low-power diodes. By contrast, SPLPL90\_3 (\(75\,\mathrm{W}\) CW) accepts substantially higher drive; with the supply at \(80\,\mathrm{V}\) it remains functional and reaches about \(100\,\mu\mathrm{J}\) per pulse at \(10\,\mathrm{kHz}\).

Changing the storage capacitance has a different effect. Larger capacitance increases the current pulse widths and  also increases energy without raising the peak voltage. Within the range we tested there is still comfortable margin, since the peak current stays the same and the duty cycle only marginally increased. Even with a longer pulse, the low-power diodes do not reach their limit, and the driver stays stable. By increasing the capacitance, we can still push the power per pulse to about four times the original setup while maintaining a sub- 100 ns pulse width. Unlike \(V_{\text{laser}}\), increasing capacitance does not increase the risk of damaging the laser diodes. Based on our analysis, the final list of components used in our circuit driver is shown in \autoref{tab:key-components}.

\begin{table}[t]
\centering
\vspace{2 mm}
\caption{Key components used in the driver design.}
\footnotesize
\begin{tabular}{c|c}
\hline
\textbf{Component} & \textbf{Value / Part} \\
\hline
R1 & 270 $\Omega$ \\ \hline
C1 & 100 nF \\ \hline
U1 & LMG3100R044VBER \\ \hline
D1 & BAS5202 \\ \hline
R2 & 0.027 $\Omega$ \\
\hline
\end{tabular}
\label{tab:key-components}
\vspace{-2mm}
\end{table}

\vspace{-1 mm}
\section{Watertank Experiment}
\label{sec:4-watertank}

We have so far established that our transmitter design trades high pulse energy for a much higher repetition rate compared to commercial systems. The crucial system-level question now is: \textit{Can these lower-energy pulses produce a detectable PA signal in a realistic setting?} We answer this using a water-tank experiment, a standard validation in PA research. Water provides a controlled, homogeneous medium with a sound speed similar to soft tissue, allowing us to test our system in a medium that mimics the human body's acoustic properties but without biological variability. This experiment allows us to directly assess whether our transmitter's high repetition rate averaging can compensate for its lower per-pulse energy.

 \begin{figure}[t]
    \centering
    \begin{subfigure}[t]{0.53\linewidth}
    \centering
    \includegraphics[width=0.95\linewidth]{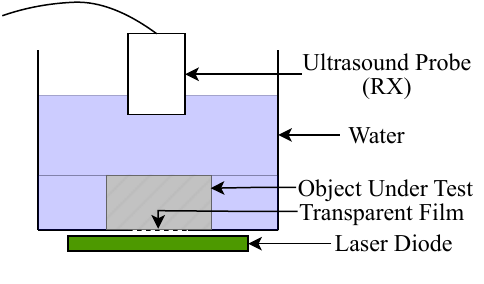}
    \caption{Diagram of experiment.} 
    \label{fig:PA-water-diagram}
    \end{subfigure} 
    \hfill
    \begin{subfigure}[t]{0.4\linewidth}
    \centering
    \includegraphics[width=0.95\linewidth]{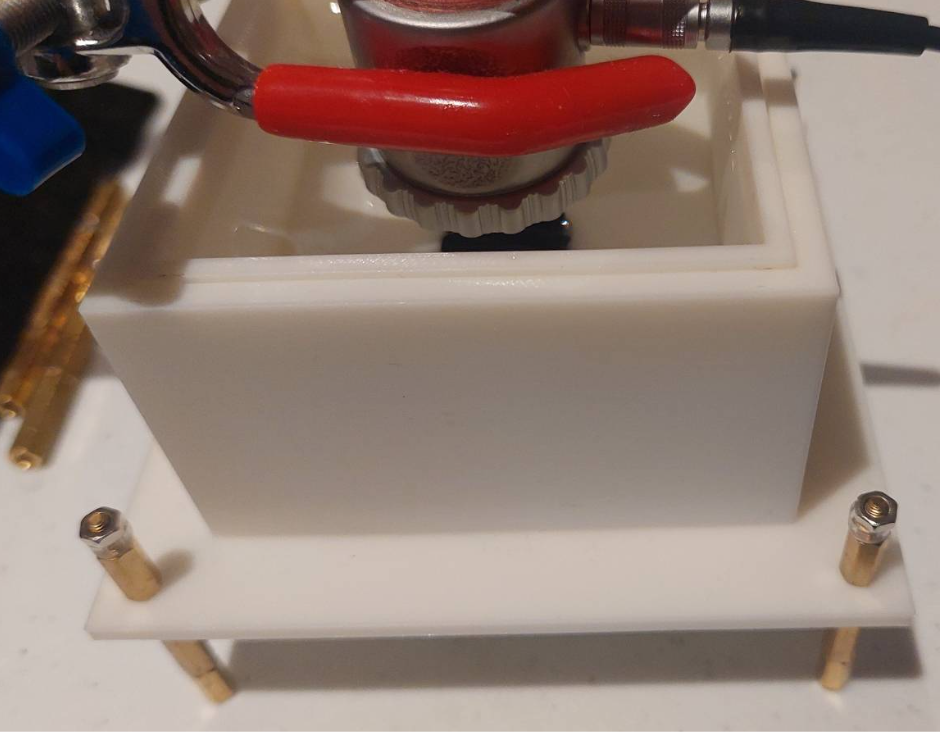} 
    \caption{Setup of experiment.}
    \label{fig:PA-water-setup}
    \end{subfigure} 
\caption{Watertank experiment diagram and setup.}
\label{fig:PA-water-experiment}
\end{figure}

\begin{figure}[t]
    \centering
    \includegraphics[width=0.98\linewidth]{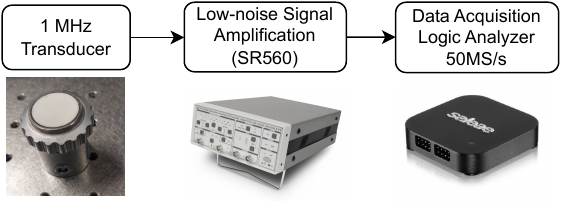}
    \vspace{-2mm}
    \caption{Receiver chain for the PA signals.}
    \vspace{-2mm}
    \label{fig:PA-receiver-bench-instrument}
\end{figure}


\vspace{-2mm}
\subsection{Experiment Setup}

We use a standard water-tank transmission setup, where the transmitter (laser diode) and the receiver (ultrasound probe) are on opposite sides of the sample, as shown in \autoref{fig:PA-water-diagram}. From the bottom of the setup, a laser diode emits light through a transparent film into a black rubber target (object under test) and a water bath. The light absorbed by the target (rubber) generates a PA pulse that travels through the water and is detected by a 1 MHz, 20 mm-diameter ultrasound probe positioned above.
This arrangement is used as a first test in PA sensing for several reasons. The black rubber target acts as a simple, high-contrast substitute for biological light absorbers, providing a strong signal to confirm the system's basic functionality. Water has a much lower optical absorption than the rubber, ensuring that light is absorbed primarily by the target. Furthermore, the known speed of sound in water (1{,}480 m/s) allows for simple time-of-flight verification of the PA signal, as we show next.

\begin{figure}[t]
    \centering
    \includegraphics[width=0.7\linewidth]{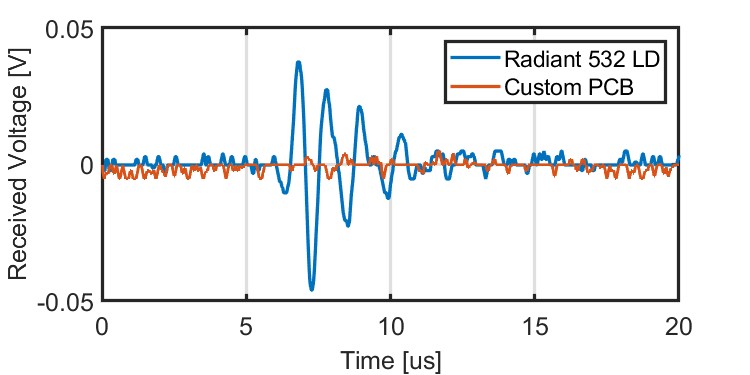}
    \vspace{-1mm}
    \caption{Received voltage at a single pulse}
    \vspace{-1 mm}
    \label{fig:received-signal-single-pulse}
\end{figure}

\begin{figure*}[t]
    \centering
    \begin{subfigure}[b]{0.3\linewidth}
        \centering
        \includegraphics[width=1.1\linewidth]{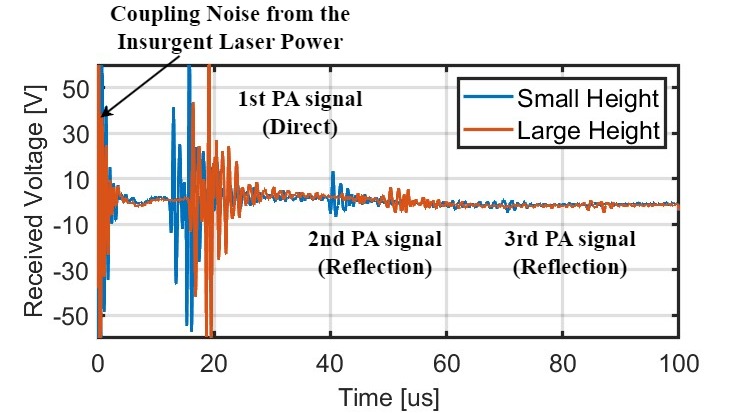}
        \caption{PA signals from our design.}
        \label{fig:pa-signal-ld}
    \end{subfigure}
    \hfill
    \begin{subfigure}[b]{0.3\linewidth}
        \centering
        \includegraphics[width=1.1\linewidth]{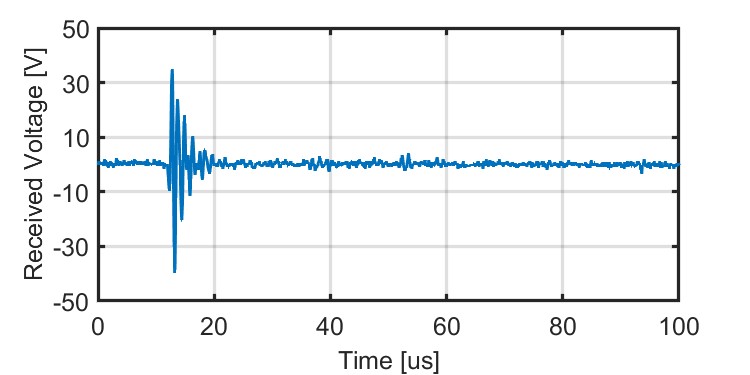}
        \caption{PA signal from commercial system}
        \label{fig:pa-signal-radiant}
    \end{subfigure}
    \hfill
    \begin{subfigure}[b]{0.3\linewidth}  
        \centering
        \includegraphics[width=1.1\linewidth]{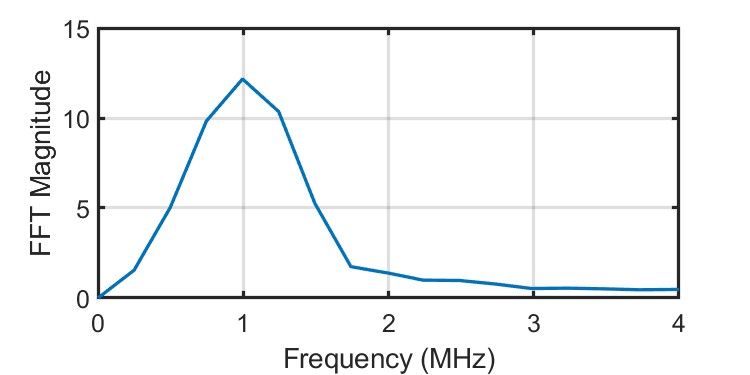}
        \caption{FFT of PA signal from laser diode}
        \label{fig:pa-signal-fft}
    \end{subfigure}
\caption{PA signals from custom and commercial systems.}
\label{fig:pa-signal-rubber-all}
\end{figure*}

We evaluate detectability at two probe–target separations: a \emph{small height} ($\sim$2 cm) and a \emph{large height} ($\sim$2.5 cm). The small height establishes a baseline PA signal and anchors timing; the large height lengthens the path, reducing SNR and delaying arrivals, which also emulates monitoring shallower vs.\ deeper on-body tissues.
The signal recorded by the ultrasound probe then passes through a low-noise preamplifier (SR560) with 40 dB gain. A logic analyzer records the amplified waveforms continuously for one second, the recording is then segmented into individual PA signals and summed across segments for analysis. To compare our system vs. a ground truth, we also repeat this experiment using the commercial PA system, OPOTEK Radiant 532 LD. 

\vspace{-2mm}
\subsection{Detecting a PA signal}

As expected from the transmitter characterization in \autoref{sec:3-transmitter}, the energy from a \textit{single pulse} of our laser diode is insufficient to generate a signal visible above the noise floor (red curve in \autoref{fig:received-signal-single-pulse}), while a single high-energy pulse from the commercial system provides a reliable PA signal (blue curve in \autoref{fig:received-signal-single-pulse}). 
However, the strength of our system is in the high and stable repetition rate provided by our design. \autoref{fig:pa-signal-ld} shows that, by firing the laser at 10 kHz and averaging the received signals over a one-second window (10,000 pulses), we recover a clear PA signal with an SNR comparable to that of the commercial system, which fires only 10 high-energy pulses in the same interval (\autoref{fig:pa-signal-radiant}). This result confirms that our design choice, exchanging high pulse energy for a high repetition rate, is a viable strategy for generating detectable PA signals with low-cost, low-power, compact hardware.

An extra test required to validate that we are measuring a true PA signal is to verify its timing against the physical setup. With approximately 2 cm between the target and the probe, we expect the first arrival to appear at $\sim$13\,\textmu s after the trigger. This matches the peak we see in the blue curve in \autoref{fig:pa-signal-ld}. A second peak at $\sim$40\,\textmu s matches a longer reflected path about three times the direct distance. In addition, a genuine acoustic signal should take longer to arrive at a farther distance, while electrical interference would appear at the same time. As shown in \autoref{fig:pa-signal-ld}, the coupling noise for both distances occurs at the same time, while the signal arrival is delayed for the larger height, confirming its acoustic nature. Finally, in the frequency domain, the FFT concentrates near 1\,MHz, consistent with the center frequency of the ultrasound probe (\autoref{fig:pa-signal-fft}). Together, these checks support that the recorded waveforms are PA pulses with properties similar to those of a commercial system. 

\vspace{-1 mm}
\section{Embedded Receiver Design}
\label{sec:5-receiver}

In the previous sections, we show that our custom transmitter, despite limited power, produces a clear PA pulse measured with advanced (benchtop) receiver instruments (\autoref{fig:PA-receiver-bench-instrument}). In this section, we propose a compact receiver design. The receiver chain of our platform is simpler than the transmitter, as there is no high-current switching or nanosecond pulse requirements, but it is still not trivial to implement on a small embedded platform: \textit{Individual PA signals are extremely faint and sits near the noise floor, requiring careful amplification and high-speed digital processing to be recovered.} Our goal is to replicate the performance of the benchtop receiver and processing chain on a small, portable board, enabling a full PA sensing system (transmitter and receiver) that can work outside of a laboratory.

\begin{figure*}[t]
    \centering
    \begin{subfigure}[t]{0.31\linewidth}
        \centering
        \includegraphics[width=0.9\linewidth]{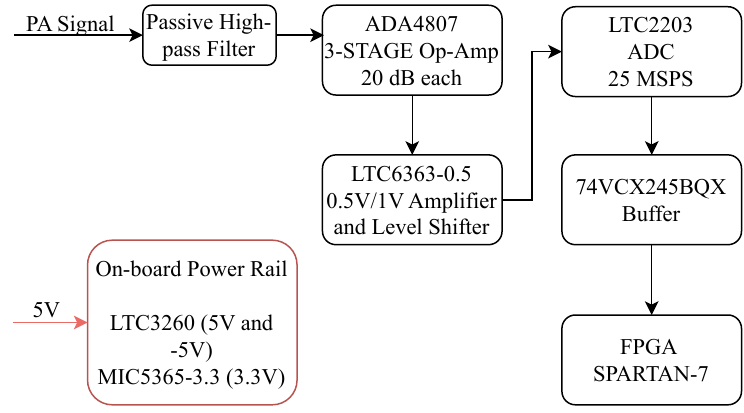}
        \caption{Main components.}
        \label{fig:receiver-flow-chart}
    \end{subfigure}
    \hfill
    \begin{subfigure}[t]{0.31\linewidth}  
        \centering
        \includegraphics[width=0.9\linewidth]{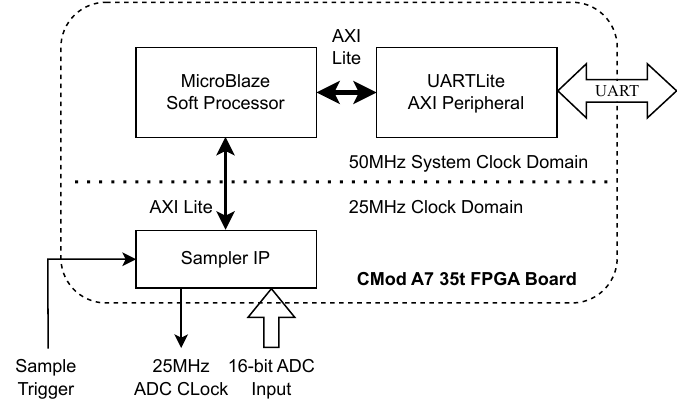}
        \caption{FPGA system diagram.}
        \label{fig:fpga-flow-chart}
    \end{subfigure}
    \hfill
    \begin{subfigure}[t]{0.35\linewidth}
        \centering
        \includegraphics[width=0.9\linewidth]{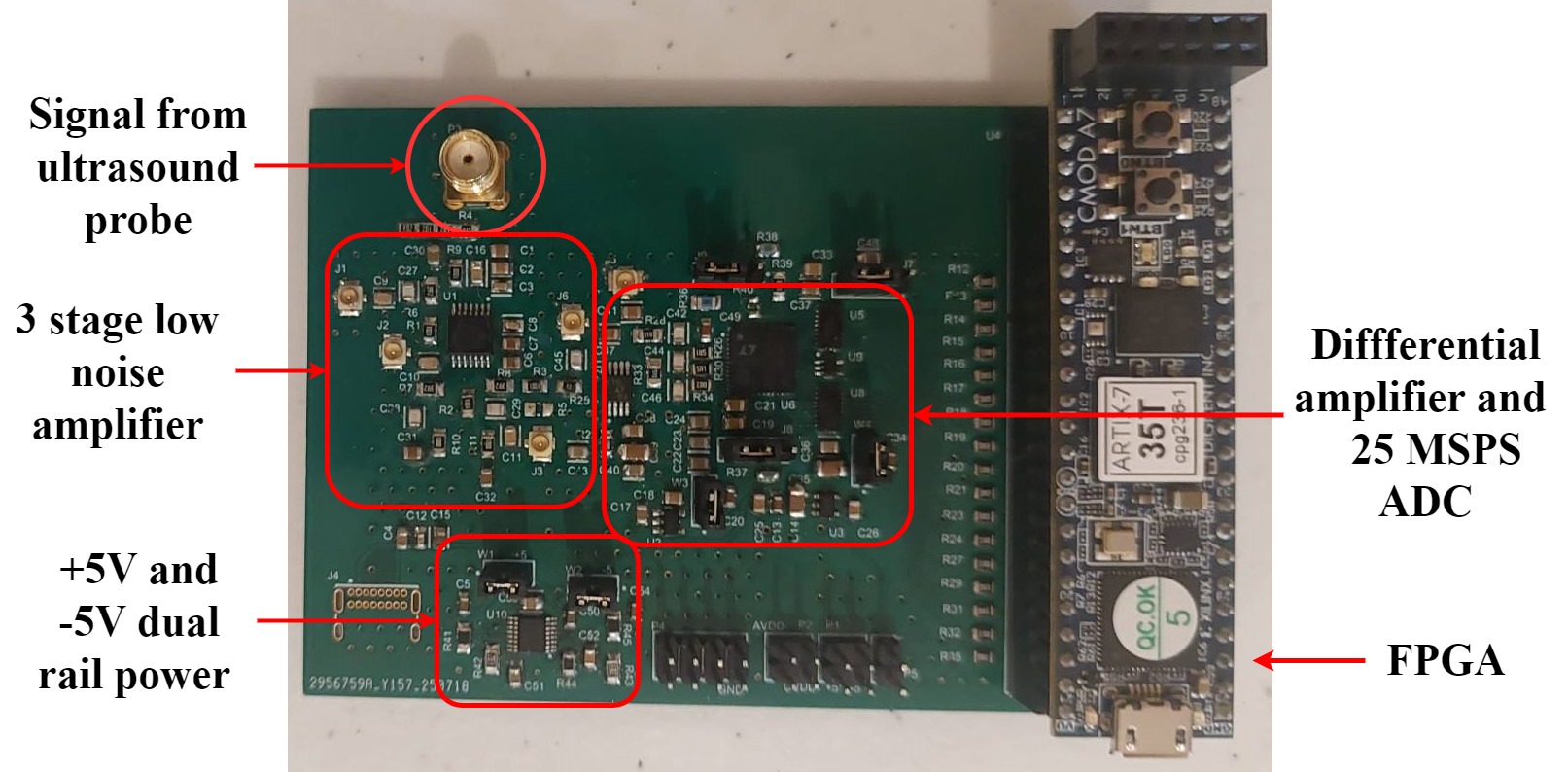 }
        \caption{PA receiver board.}
        \label{fig:PA-receiver-board}
    \end{subfigure}
\caption{Receiver board design.}
\label{fig:receiver-board-design}
\end{figure*}

\vspace{-3 mm}
\subsection{Receiver Design}

Our receiver has three stages, each addressing a specific challenge in recovering the faint PA signals. The main components of the system are shown in \autoref{fig:receiver-flow-chart}. The first stage (amplification) is required to enhance the low-energy pulse sent by our transmitter. The second and third stages are needed to overlap and average the individual pulses into a decodable signal. This tight co-design of transmitter and receiver is critical to obtain a compact PA system.   

\textit{Stage 1: Amplifying the faint analog signal.} The raw signal from the ultrasound transducer is on the order of $\mu$V, far too weak to be digitized directly and can be easily lost in electronic noise. To address the low SNR, we design a high-gain, low-noise amplification stage. 
Our implementation uses a passive high-pass filter to remove slow drift, followed by three cascaded ADA4807 op-amp stages (60 dB total). We use a multi-stage design because it preserves signal bandwidth better than a single high-gain stage. An LTC6363 amplifier provides the final conditioning before digitization. 

\textit{Stage 2: Digitizing the high-frequency waveform.} Converting a high-speed signal into a digital form is an inherently challenging process, as high-speed signals are more susceptible to electronic noise and timing errors. We use high-speed sampling to create a high-resolution digital snapshot of the waveform. This is accomplished with an LTC2203 Analog-to-Digital Converter (ADC) running at 25 MSPS. This high sample rate is much faster than the signal's 1 MHz center frequency, ensuring we accurately capture its features. A 74VCX245 buffer isolates the ADC's fast outputs to maintain signal integrity during the transfer for processing.

\textit{Stage 3: Extracting the signal from noise.} Even after amplification, the signal from a single pulse has extremely low SNR. The key is to leverage the thousands of pulses generated by our high-repetition-rate transmitter. For this, we perform real-time averaging directly on the FPGA board (\autoref{fig:fpga-flow-chart}). For each pulse, it captures a 200 $\mu$s window of ADC data and adds it to an accumulating buffer in memory. This process amplifies the PA signal while the random noise averages out toward zero. A MicroBlaze soft processor handles configuration, allowing a user to read the final, clean waveform over a simple UART link. This on-device processing makes the receiver a truly self-contained, embedded system.

\subsection{Receiver Prototype}

To validate our design, we built a compact prototype receiver (60 mm × 100 mm) and performed an end-to-end system test (\autoref{fig:PA-receiver-board}). 
The goal is to demonstrate that our complete embedded platform, which combines the low-energy transmitter from \autoref{sec:3-transmitter} with the new receiving and processing chain, can successfully reproduce the results of the watertank experiment from \autoref{sec:4-watertank} without requiring any external benchtop equipment.

\begin{figure}[t]
    \centering
    \includegraphics[width=0.85\linewidth]{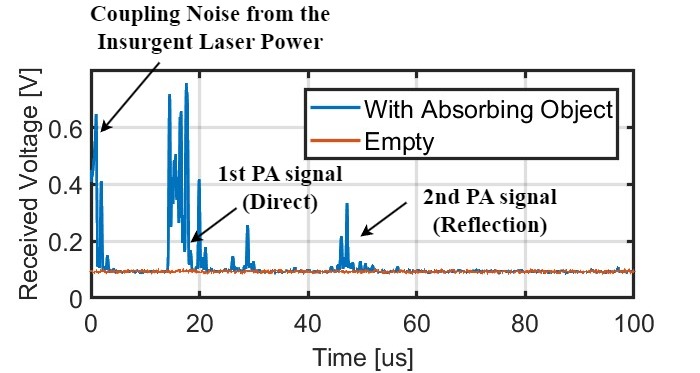}
    \caption{PA signals recorded with the receiver board. } 
    \vspace{-3mm}
    \label{fig:custom-receiveri-output}
\end{figure}

As shown in \autoref{fig:custom-receiveri-output}, the fully embedded system successfully recovered the same faint PA signals, with clear direct and reflected arrivals consistent with the bench measurements. 
The appearance of the waveform differs from the signals in \autoref{fig:pa-signal-rubber-all} because the unipolar ADC records only positive values, so negative voltages are effectively “folded” into the positive range (similar to rectification).
%
These effects, however, do not change the processing. 
This is a crucial result: it confirms that \textit{our on-board amplification and real-time averaging are effective enough to detect the weak pulses generated by our custom transmitter}. This demonstrates a functional, self-contained PA system from laser light emission to the final digital PA signals.

The transmitter draws around 305 mW , and the receiver consumes around 890 mW during active sampling. This reflects the use of high-speed components needed for multi-MHz sampling (ADC) and on-device real-time processing (FPGA). This puts the total system power budget at 1.2 W. On a standard 3{,}000 mAh USB power bank, this would allow for over 4 hours of continuous operation for the entire platform. In a practical, wearable scenario, the system would be duty-cycled, sleeping between measurements, which could extend the battery life to many days.

\vspace{-1 mm}
\begin{figure*}[t]
    \centering
    \begin{subfigure}[t]{0.23\linewidth}
        \centering
        \includegraphics[width=\linewidth]{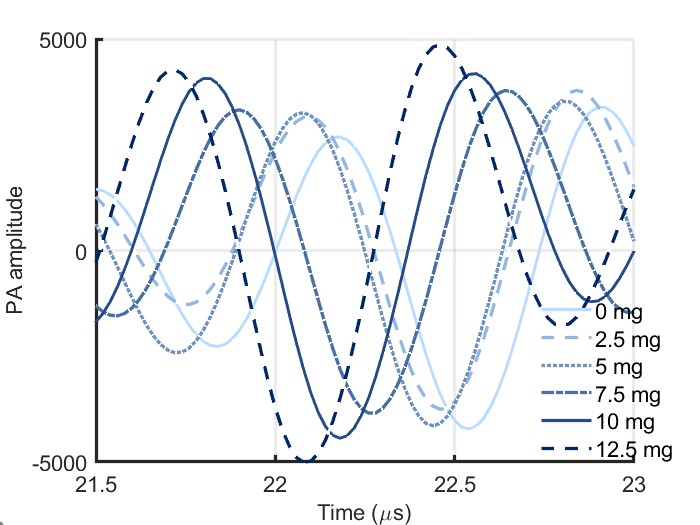}
        \caption{PA signals from glucose water solutions.}
        \label{fig:pa-glucose-solution-experiment-results}
    \end{subfigure}
    \hfill
    \begin{subfigure}[t]{0.23\linewidth}
        \centering
        \includegraphics[width=\linewidth]{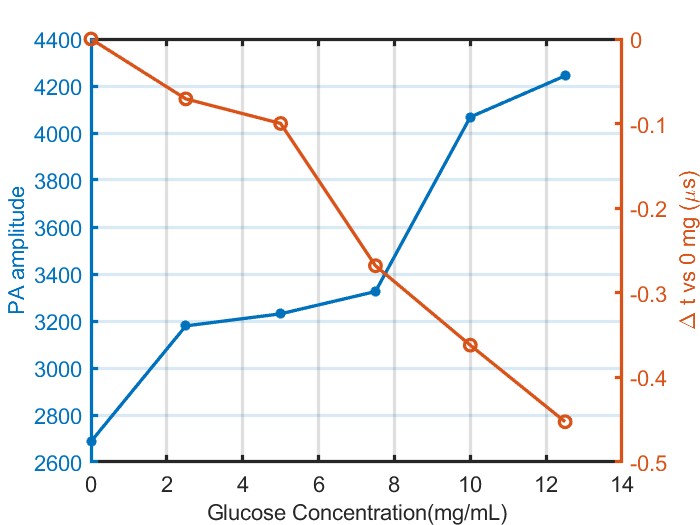}
        \caption{Glucose arrival time shift and amplitude variations with concentration.}
        \label{fig:pa-glucose-time-shift}
    \end{subfigure}
    \hfill
    \begin{subfigure}[t]{0.23\linewidth}
        \centering
        \includegraphics[width=\linewidth]{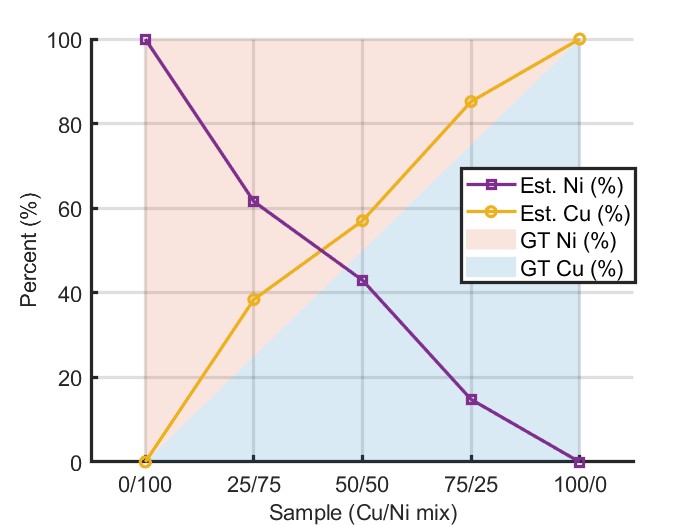}
        \caption{CuSO\textsubscript{4}/NiSO\textsubscript{4} unmixing results. GT represents ground truth results.}
        \label{fig:Cu-Ni-unmixing}
    \end{subfigure}
    \hfill
    \begin{subfigure}[t]{0.25\linewidth}
    \centering
    \includegraphics[width=0.7\linewidth]{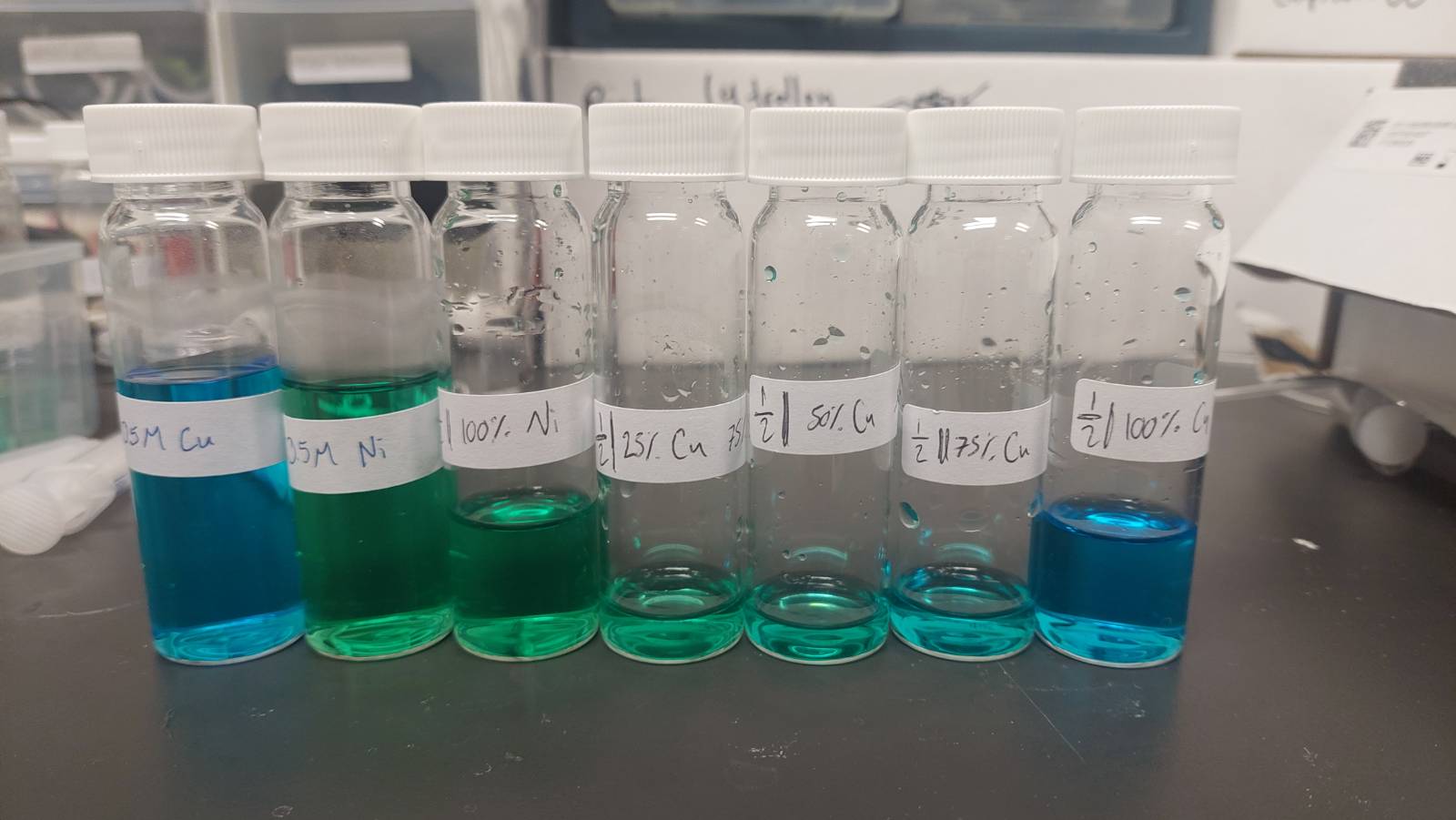}
    \caption{CuSO\textsubscript{4} and NiSO\textsubscript{4} solutions in different combinations.}
    \vspace{-3mm}
    \label{fig:cu-ni-solution}
    \end{subfigure}
    \caption{Experimental results of glucose solution and spectral unmixing.}
    \label{fig:experiment-results}
\end{figure*}

\section{Experimental demonstration}
\label{sec:6-evaluation}

In this section, we take a first step toward multi-biomarker PA sensing with two controlled chemical demonstrations: (1) tracking glucose concentration in water and (2) unmixing copper and nickel sulfates solutions, which are commonly used as stable, simpler solutions that serve as a reliable stand-in for the more complex biological molecules of oxygenated and deoxygenated hemoglobin \cite{Fonseca2017Sulfates, Fonseca2017ThreeDPA}.
We choose these biomarkers because they are central for clinical monitoring and treatment \cite{Ghazaryan2018ExtendedNIR, IDFAtlas11_2025}. 
The ability to monitor changes in glucose concentration is foundational for non-invasive diabetes management. The second demonstration, differentiating between oxy- and deoxy- hemoglobin, is the principle behind measuring blood oxygen saturation (\ce{SpO2}). This is a critical vital sign used to monitor vascular and respiratory health. These two tests, therefore, serve as foundational proofs-of-concept for some of the most impactful potential applications of this technology.


%

%
%

\vspace{-3 mm}
\subsection{Glucose Solution}

The primary requirement for any non-invasive diabetes monitor is the ability to reliably track whether glucose levels are rising or falling. This experiment tests that core function. By preparing solutions with increasing glucose concentrations, we validate a fundamental question: does our system produce a consistent, predictable signal change that corresponds to different glucose levels? Proving this capability is an essential first step toward building a device that can one day distinguish between low, normal, and high glucose states in the body in a non-invasive manner.

Under a setup with fixed optics and geometry (\autoref{fig:PA-water-experiment}), we prepare glucose solutions from 0 to 12.5 mg/mL. This range partially overlaps with physiological values. Typical blood glucose is $\sim$0.7 to 1.4 mg/mL, and people with diabetes can exceed 2 mg/mL \cite{ADA2025_DiagnosisClassification}. The higher end of our range also aligns with nonclinical settings: fruit juices have 10 to 50 mg/mL (or even higher for certain fruits) \cite{Li2020}. Overall, this range lets us test well-defined levels while staying relevant to both medical and nonclinical contexts. 

For our analysis, each trace is cross-correlated with a reference (water with no glucose). To estimate changes in glucose levels, it is central to estimate the arrival time and peak amplitude of the PA signals. The received traces are shown in \autoref{fig:pa-glucose-solution-experiment-results}; and \autoref{fig:pa-glucose-time-shift} summarizes both trends: as concentration increases, the arrival time shifts earlier and the peak amplitude increases across the range.
Both trends are consistent with underlying glucose physics. The earlier arrival time corresponds to a known increase in the speed of sound as glucose concentration rises. More importantly for this application, the increase in signal amplitude is the direct result of higher glucose concentrations absorbing more light, producing a stronger PA response. These results show that our system detects the critical property of concentration shifts in arrival time and amplitude. 

\vspace{-2mm}
\subsection{Spectral Unmixing}

We now tackle a more complex challenge that directly mimics the real-world application of measuring blood oxygen saturation (\ce{SpO2}). This measurement requires a device to distinguish between two different forms of hemoglobin (oxygenated and deoxygenated) that are always mixed together in the blood.
This chemical mixture behaves linearly: the total PA signal we measure is simply the sum of the individual signals from the copper and nickel components, weighted by their concentration. 
%
%
Because this is the same fundamental challenge as separating the signals from two types of hemoglobin in blood, this test is directly relevant for validating our system's potential for blood-oxygen measurements \cite{Fonseca2017Sulfates,Fonseca2017ThreeDPA}.
We use the 638 nm and 905 nm laser diodes because at these specific wavelengths, the absorption coefficients of copper and nickel sulfate are significantly different, providing a distinct signature for each chemical. 

For each sample in \autoref{fig:cu-ni-solution}, the PA amplitudes, $P(\lambda)$, were normalized by the corresponding per-pulse laser energy, $E(\lambda)$ (which we measured in \autoref{sec:3-transmitter}). The resulting energy-normalized measurement vector, $\mathbf{B}$, is modeled as a linear function of the molar absorption coefficients ($\epsilon$) and effective concentrations ($C$) of copper and nickel:
\begin{equation}
    \mathbf{B} = 
    \begin{bmatrix} P(\lambda_1) / E(\lambda_1) \\ P(\lambda_2) / E(\lambda_2) \end{bmatrix} 
    = 
    \begin{bmatrix} 
    \epsilon_{Cu}(\lambda_1) & \epsilon_{Ni}(\lambda_1) \\ 
    \epsilon_{Cu}(\lambda_2) & \epsilon_{Ni}(\lambda_2) 
    \end{bmatrix} 
    \begin{bmatrix} C_{Cu} \\ C_{Ni} \end{bmatrix}.
\end{equation}

This equation can be expressed in matrix form as $\mathbf{B} = \mathbf{M} \mathbf{C}$, where $\mathbf{M}$ is the $2 \times 2$ extinction matrix containing the known absorption coefficients (from the SIMPA library \cite{Groehl2022SIMPA}) and $\mathbf{C}$ is the unknown concentration vector. The concentrations for each sample are found by solving this system of linear equations: $\mathbf{C} = \mathbf{M}^{-1} \mathbf{B}$. Once the effective concentrations $C_{Cu}$ and $C_{Ni}$ are determined, the raw fractional concentration of copper, $f_{\text{raw}}$, is calculated for each sample:

\begin{equation}
    f_{\text{raw}} = \frac{C_{Cu}}{C_{Cu} + C_{Ni}}.
\end{equation}

To align the measurements with the experimental design, where samples spanned the full range from 0\% to 100\% copper, the raw fractions were scaled using an endpoint-anchored affine transformation. This normalization maps the calculated fraction of the first sample ($i=1$) to 0 and that of the final sample ($i=N$) to 1, thereby correcting for baseline offsets or scaling discrepancies.


%
\autoref{fig:Cu-Ni-unmixing} compares the estimated composition against the intended mixture for five Cu/Ni samples. The shaded backgrounds indicate the ground truth (GT): blue for Cu\% and orange for Ni\%. The purple and yellow curves show the estimated Ni\% and Cu\% from the unmixing performed with received PA signals. The plot suggests that two-wavelength unmixing tracks the intended mixture proportions.

Together, the glucose and Cu/Ni sulfate tests indicate that our platform is able to detect concentration-dependent shifts in arrival time and amplitude and can recover binary mixture proportions with two wavelengths. These outcomes align with the design goals and offer a promising basis for extending the platform to additional wavelengths and to tissue-mimicking or more complex environments.

\vspace{-3 mm}
\section{Related Work}

We review three areas related to embedded PA sensing. \textit{PA imaging} uses the same underlying mechanism and demonstrates a wide range of applications. \textit{PA sensing} focuses on concentration and composition, aiming at quantity-oriented outputs. Work on \textit{embedded sensing} shows common approaches for small, low-power systems. Together, these areas motivate our design choices and the scope of our study.

\noindent\underline{\emph{PA imaging.}}
PA imaging has demonstrated broad clinical potential, with studies in dermatology \cite{Oh2006}, oncology \cite{Ermilov2009}, rheumatology \cite{Xu2013}, neuroscience \cite{Na2022}, and ophthalmology \cite{Jiao2010}, and with two commercial systems receiving FDA clearance and CE marking \cite{Riksen2023, Manohar2016}.
These clinical applications underscore the relevance of PA imaging while also revealing practical limitations. Today’s PA imaging systems often rely on high-power lasers and high-channel ultrasound hardware, which drives significant cost, size, and workflow complexity. Active research on light sources and detectors aims to improve detection sensitivity, portability, cost, and footprint. For example, by using LEDs instead of lasers \cite{Singh2020LEDbasedPA, Joseph2021LEDdualPAUS} or overdriving low-cost laser diodes \cite{Stylogiannis2018, Seeger2023}, PA imaging systems could move from specialized labs into point-of-care settings. Nonetheless, little has been demonstrated using embedded systems to date. Our simplified system reduces complexity and cost while maintaining focus on potential applications.

\noindent\underline{\emph{PA sensing.}} 
PA sensing (sometimes also called PA spectroscopy) is used across domains, including gas and breath analysis \cite{Palzer2020PhotoacousticGasSensingReview}, food and agriculture \cite{Fiorani2021, Wali2024}, materials characterization \cite{Isaiev2022NanomaterialsPA}, and health monitoring \cite{Jin2022PASHealth}. However, similar to PA imaging, these devices rely on high-power pulsed lasers.
Few studies have explored compact PA sensing systems, but they still rely on bulky and costly benchtop hardware \cite{Liu2018, Guo2024PortablePApH}.
Two works are particularly relevant. A wrist-worn PA sensor reports human data on superficial blood vessels and blood-flow changes, which demonstrates that on-body operation is possible~\cite{Zhang2024WearablePAWatch}. However, the laser, optics, electronics, and power supply are housed in a backpack (7 kg) that is connected to the wrist unit by tethered cables. The backpack still needs to be connected to an external power supply. A second work, \emph{Nutrilyzer}, builds a low-cost PA spectrometer from LEDs to quantify liquid samples in a small cuvette \cite{Rahman2016Nutrilyzer}. This represents a step toward portable PA sensing, but it remains a benchtop system and uses 10 Hz LED light with a 50\% duty cycle, which is not strictly a PA signal. Together, the above studies point towards the need for a low-power, embedded PA system that packages the light source, acoustics, and processing in a single, deployable device.

\noindent\underline{\emph{Sensing with other modalities.}} 
The systems community has built biomedical sensors using light, acoustics, and RF based modalities. 
Optical approaches (cameras, LEDsa, and photodiode wearables) are compact, but they mainly measure shallow blood-flow changes at the skin and are sensitive to motion and ambient light \cite{Castaneda2018WearablePPGReview}.
Phone acoustics can screen sleep apnea or estimate lung function with speaker–microphone hardware, yet they sense motion or airflow rather than chemistry and often require controlled posture \cite{Nandakumar2015ApneaApp,Song2020SpiroSonic}.
RF systems recover breathing and sometimes heart rate through clothing and walls, but performance degrades with multipath and movement, and the signals reflect bulk dielectric properties, not molecular content \cite{Hillyard2018XTRFResp,Shang2016WiFiVitalSigns}.
%

\vspace{-3 mm}
\section{Discussion}

This work delivers a first prototype that breaks a critical initial barrier for \textit{embedded} PA sensing: an end-to-end optoacoustic–processing chain that, using compact hardware, reliably detects signals and spectrally distinguishes targets under controlled conditions. Testing in homogeneous media (water) and solution phantoms (glucose and Cu/Ni mixtures) isolates the sensing pathway, allowing measured changes to be attributed to the chain rather than to tissue variability or motion. These experiments demonstrate stable PA arrivals and consistent two-wavelength discrimination in simple solutions, establishing the foundation for deployment in heterogeneous tissues and opening the path toward quantitative, on-body sensing. Despite these achievements, a few barriers remain before embedded PA can be a reality.

\noindent\underline{{Barrier 1: Robustness in dynamic media.}} Water provides a homogeneous media to isolate the sensing chain and verify end-to-end behavior. The next step is to carry out experiments in more complex media such as fruit juice, along with tissue-mimicking phantoms that approximate skin, fat, and muscle. More complex media will help reveal how various layers affect the PA signals. Running experiments in these settings will both confirm stability and guide targeted improvements, so the system remains reliable outside a controlled environment. These studies are the natural next steps for this work and move the prototype from proof-of-principle toward practical sensing.

\noindent\underline{{Barrier 2: Absolute quantity readouts.}} We are able to show concentration-dependent changes in glucose solution and Cu/Ni sulfate mixtures, evidence that the system captures meaningful, spectrally interpretable variation. Building on these relative results, the next step is to turn trends into predictive conditions. One such example is predictive modeling, using information across wavelengths and time to classify or predict conditions (for example, flagging high glucose ranges). This type of prediction will bring our platform closer to user-ready devices.

\noindent\underline{Barrier 3: An improved ultrasound receiver.}
In this study, we used an off-the-shelf, single-element resonant receiver to keep the system simple. The trade-off is limited bandwidth and minimal spatial information. Spatial information and broad bandwidth are key in many applications: it enables source localization, separation from background, rejection of reflections and motion, and stable performance when placement or coupling changes. Multiple-receiver arrays would enable more predictions for a wider range of applications.

\vspace{-2 mm}
\section{Conclusion}


This work establishes the first step toward embedded PA. 
Guided by transmitter–receiver co-design insights, we explore the use of low-cost, low-power laser diodes as compact, high-repetition optical sources. 
Building on these insights, we perform evaluations in controlled environments that show concentration-dependent trends for glucose and two-wavelength unmixing of Cu/Ni sulfate mixtures. These results demonstrate the potentials for multi-biomarker sensing.  
With this foundation, the remaining challenges to bring embedded PA closer to the user are enhanced operation in everyday and layered media, quantitative or predictive recommendations, and receivers that add spatial information. 
Overall, our findings mark a first step and widen the design space for low-cost, portable PA sensing.

\newpage
\bibliographystyle{ACM-Reference-Format}
\bibliography{acmart}

\end{document}